\documentclass{article}

\usepackage{arxiv}

\usepackage[utf8]{inputenc} % allow utf-8 input
\usepackage[T1]{fontenc}    % use 8-bit T1 fonts
\usepackage{hyperref}       % hyperlinks
\usepackage{url}            % simple URL typesetting
\usepackage{booktabs}       % professional-quality tables
\usepackage{amsfonts}       % blackboard math symbols
\usepackage{nicefrac}       % compact symbols for 1/2, etc.
\usepackage{microtype}      % microtypography
\usepackage{lipsum}
\usepackage{graphicx}

\title{Handling Position Bias for Unbiased Learning to Rank in Hotels Search}

\author{
  Yinxiao Li \\
  TripAdvisor, Inc.\\
  400 1st Avenue\\
  Needham, MA 02494 \\
  \texttt{liyinxiao1227@gmail.com} \\
}

\begin{document}
\maketitle

\begin{abstract}
Nowadays, search ranking and recommendation systems rely on a lot of data to train machine learning models such as Learning-to-Rank (LTR) models to rank results for a given query, and implicit user feedbacks (e.g. click data) have become the dominant source of data collection due to its abundance and low cost, especially for major Internet companies. However, a drawback of this data collection approach is the data could be highly biased, and one of the most significant biases is the position bias, where users are biased towards clicking on higher ranked results. In this work, we will investigate the marginal importance of properly handling the position bias in an online test environment in Tripadvisor Hotels search. We propose an empirically effective method of handling the position bias that fully leverages the user action data. We take advantage of the fact that when user clicks a result, he has almost certainly observed all the results above, and the propensities of the results below the clicked result will be estimated by a simple but effective position bias model. The online A/B test results show that this method leads to an improved search ranking model. 

\end{abstract}

% keywords can be removed

\section{Introduction}

With an increasing presence of machine learning in search ranking and recommendation systems, there is a larger demand for data than ever before. Implicit user feedbacks, such as clicks, conversion and dwell time \cite{yi2014beyond}, are cheap and abundant compared with explicit human judgements, especially for large Internet companies. Thus, it has become the dominant source of data collection to solve/improve search ranking problems. However, a well-known challenge of implicit user feedbacks is its inherent bias.

Implicit user feedbacks typically include three types of bias: position bias \cite{joachims2005accurately, o2006modeling, wang2018position}, presentation bias \cite{yue2010beyond} and quality-of-context bias \cite{joachims2007evaluating}. Position bias is that users are biased towards clicking on higher ranked results, either due to laziness \cite{joachims2005accurately, wang2018position} or due to trust to the search site (trust bias) \cite{o2006modeling, joachims2007evaluating, agarwal2019addressing}. Presentation bias describes an effect where users tend to click on results with seemingly more attractive summaries, which inflates the perceived relevance. The quality-of-context bias refers to the fact that users make click decisions not only by the relevance of the clicked result, but also by the overall quality of the results in the list. Among the three biases, the position bias has the strongest effect on what users click \cite{joachims2017unbiased}. Therefore, there is a stronger demand to debias it in order to fully leverage the power of implicit user feedback data. 

\subsection{Existing work on handling position bias}
\label{handling position bias}
In this work, we will focus on the position bias due to laziness, where users may not have evaluated the whole list before making a click. This is especially true for long lists such as Tripadvisor Hotels search, where a maximum of 30 hotels are shown in a certain page. To tackle this type of bias and use implicit user data for model training, researchers have adopted three primary approaches of handling the position bias. 

The first approach is to keep all the results in the training data and neglect this position bias. This approach assumes that user has evaluated all the options \cite{covington2016deep, he2014practical, tagami2013ctr}, and it is acceptable for a relatively short list such as Facebook ad recommendation \cite{he2014practical}. The second approach is to only keep results up to the last result user clicked on \cite{grbovic2018real}. This approach assumes that user sequentially views the results from top to bottom, and will click the first relevant result as he is scrolling down and stop (similar to Cascade model  \cite{craswell2008experimental}). This works reasonably well for a relatively long list such as Airbnb search ranking \cite{grbovic2018real}. However, it has been argued that this approach is systematically biased and will lead to a ranking model that tends to reverse the existing order \cite{joachims2017unbiased, joachims2002optimizing}. 
The third approach is to keep all the results in the training data, but use the propensities as weights in the loss function \cite{wang2018position, joachims2017unbiased, aslanyan2018direct}. Compared with the previous two approaches, this approach aims at debiasing the training data by taking propensities into account. They have shown that this method leads to an unbiased loss function and thus an unbiased model, and referred to this framework as unbiased Learning-to-Rank \cite{wang2018position, joachims2017unbiased, aslanyan2018direct}. However, this approach has not yet fully leveraged the user feedback data (e.g. when a user clicks on result N, this user has almost certainly evaluated result 1 to result N-1). Besides, this approach requires propensity estimation, which is another challenging task. 

\subsection{Existing work on propensity estimation}
\label{Existing work on propensity estimation}
There has been numerous research on unbiased click propensity estimation, and position bias model \cite{wang2018position,richardson2007predicting} is one of the most classical methods. Position bias model assumes that the probability of a click on a given result is the product of the probability of evaluating the result and the probability of clicking the result given that it has been evaluated: 
\begin{equation}
\label{eq: position bias model}
P(C=1 | Hotel, k) = P(E=1 | k) \cdot P(R=1 | Hotel)
\end{equation}
where C represents whether a result is clicked, E represents whether a result is examined, R represents whether a result is relevant, and k is the position. This model in general requires result randomization experiments which degrade the user experience, although many efforts have been spent on minimizing this degradation effect \cite{radlinski2006minimally}.

To fully remove the degradation effect, Wang et al. proposed a method without result randomization, to estimate position bias from regular clicks \cite{wang2018position}. This method uses a regression-based Expectation Maximization (EM) algorithm to extract the position bias and result relevance simultaneously. A very similar dual learning method was also proposed by Ai et al. \cite{ai2018unbiased}. We would argue that this type of method tends to assign relevance based on how relevant a result is compared with other results at the same position k, potentially overlooking the fact that results appearing at top ranks are generally better than those appearing at the bottom. 

Later, Aslanyan et al. proposed a method to estimate click propensities without any intervention in the live search results. This method takes advantage of the fact that the same query-document pair may naturally change ranks over time in eCommerce search, and uses query-document pairs that appear at different ranks to estimate propensities \cite{aslanyan2018direct,aslanyan2019position}. 
Similarly, Agarwal et al. proposed an estimating method that also requires no intervention, which uses query-document pairs from different ranking functions \cite{agarwal2018consistent,agarwal2019estimating}. Both methods assume that a document will not change much over time and propensities are estimated based on the CTR of the same document at different positions. However, although documents in search engines are relatively static, the price of a hotel is very dynamic and is one of key factors to consider when users make click/booking decisions, which makes pair generation very difficult for Hotels search.

In this work, we propose to combine two of the exiting approaches of handling position bias (second and third approach in Section \ref{handling position bias}), for search ranking in Tripadvisor Hotels. We assume that user has evaluated all the hotels above the last hotel user clicked on, and will only calculate propensities of hotels below that. The propensities will be estimated by a position bias model and assigning the historical booking count as the ground truth of hotel relevance. We apply this method to generate the training data, which is used to train a model to serve live-site traffic. This model is compared with a control model which uses the training dataset where all hotel impressions are kept, and our results show that this model outperforms the control. 

\section{Method}
\label{sec:Method}

\subsection{Handling the position bias}

\begin{figure}[htp]
  \centering
  \includegraphics[width=6cm]{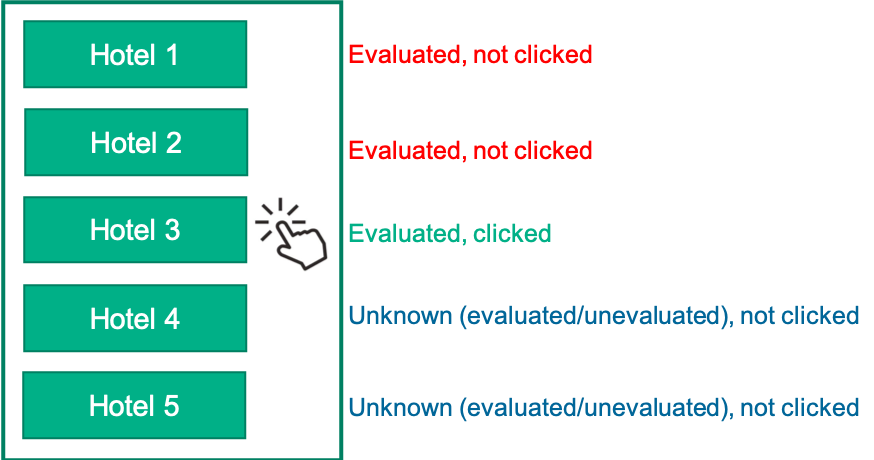}
  \caption{Concept of handling position bias.}
  \label{fig:bias_handling}
\end{figure}

Consider an example of implicit user feedback, as shown in Figure \ref{fig:bias_handling}, where there are five hotel impressions in the list, and user makes a click on Hotel 3. Here is how we will handle the position bias. Since user clicks Hotel 3, Hotel 3 is evaluated, and we assume Hotel 1-2 are also evaluated \cite{joachims2007evaluating, joachims2002optimizing}, while Hotel 4-5 are in an unknown state of being evaluated or not. So, when preparing for the training data, Hotels 1-3 will be kept while Hotel 4-5 will be sampled depending on their propensities. 

Extending the classical position bias model as shown in Equation \ref{eq: position bias model}, when user clicks on LastClickPos, user has evaluated all the results above LastClickPos:
\begin{equation}
\label{eq: propensity above}
P(E=1 | k <= LastClickedPos) = 1
\end{equation}
On the other hand, the propensity of observing results below LastClickPos can be calculated as: 
\begin{equation}
P(E=1 | k > LastClickedPos) \\= P(E=1 | k, LastClickedPos)
= \frac{P(E=1 | k)}{P(E=1 | LastClickedPos)}
\end{equation}
It is noted that this method relies on an accurate estimation of position propensity $P(E=1 | k)$. 

\subsection{Propensity estimation}
\label{sec:Propensity Estimation}

\begin{figure}[htp]
  \centering
  \includegraphics[width=10cm]{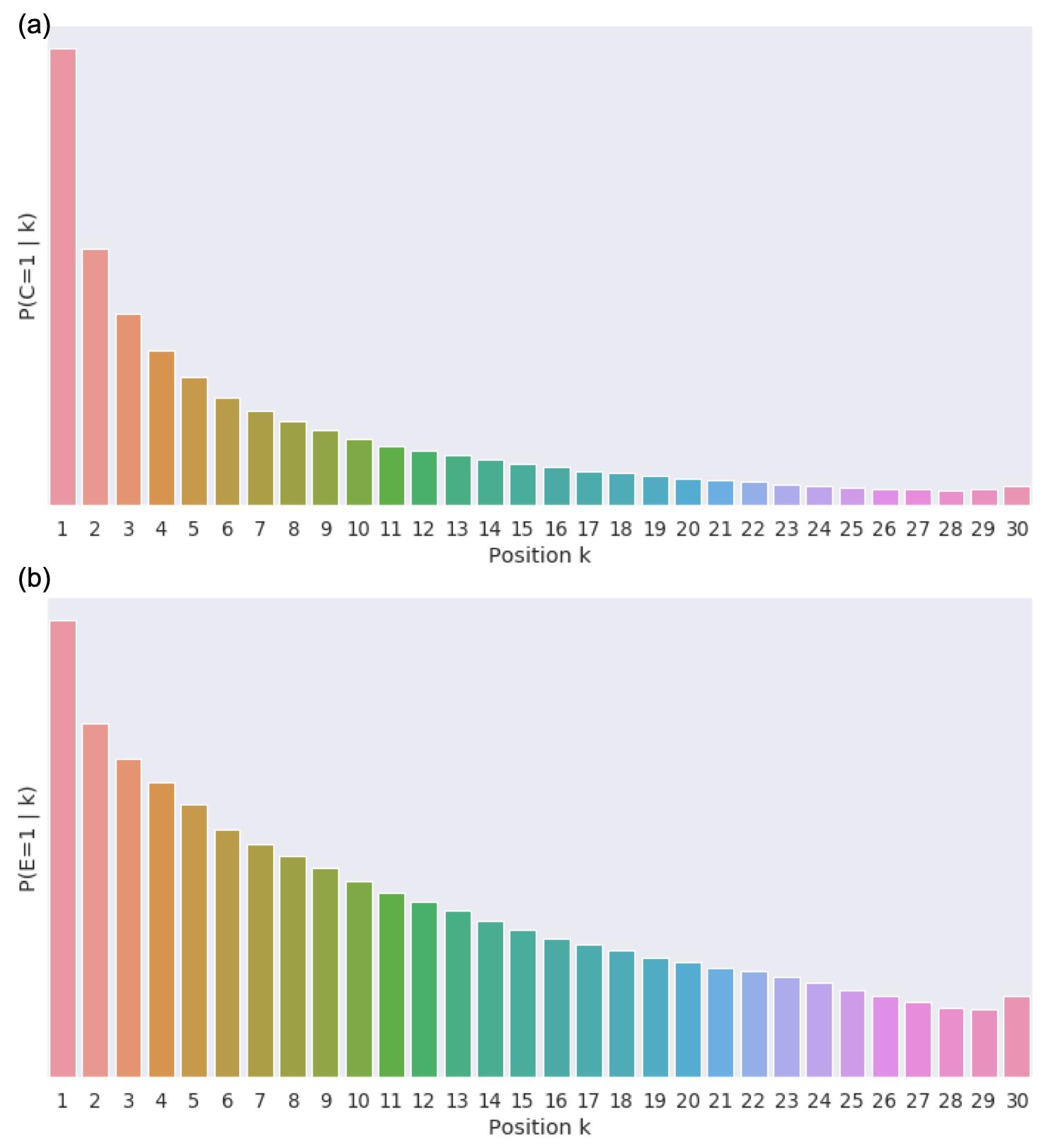}
  \caption{Click curve and propensity curve.}
  \label{fig:rel_theta}
\end{figure}

As discussed in Section \ref{Existing work on propensity estimation}, estimating propensities with the help of result randomization degrades user experience, while the existing evaluation methods from regular clicks suffer from the difficulty in separating hotel relevance from propensity. In this work, we will use a simple relevance assignment based on the historical number of bookings, which will be shown to be good enough for evaluating the average relevance of hotels on a certain position. According to the position bias model, we have:  
\begin{equation}
\label{fig:IPW-MB}
P(E=1 | k) = \frac{E[P(C=1 | k)]}{E[P(R=1 | Hotel, k)]}
\end{equation}
where we let $P(R=1 | Hotel, k)$ = historical number of bookings at position k. We have found out that for online travel agencies (OTAs), the historical number of bookings is a very strong signal on hotel relevance and is aligned with our final business goal. Figure \ref{fig:rel_theta} shows the measured click curve ($P(C=1 | k)$ vs position) and the calculated propensity curve ($P(E=1 | k)$ vs position) based on Equation \ref{fig:IPW-MB}. The click curve proves that users are highly biased towards clicking on higher ranked hotels, and since click curve is steeper than the calculated propensity curve, it means our list is already promoting more relevant hotels to the top of the list. 

\section{Experiments}
\subsection{Model}

We use LGBMRanker from lightgbm library \cite{ke2017lightgbm} to train a pairwise GBDT model. In industry, both pointwise formulation \cite{covington2016deep,ma2018entire,zhou2018deep,naumov2019deep,liu2017related} and pairwise formulation \cite{grbovic2018real,liu2017related,haldar2019applying} are widely used to solve search ranking problems, but in this work we will use pairwise methods due to its three advantages over pointwise methods: (a) Focus on learning relevant things: for example, hotels in Boston have higher CTR than hotels in Maine. A pointwise model is supposed to predict such details correctly, which is unnecessary since we will never compare a hotel in Boston with a hotel in Maine, while pairwise learning will focus on solving problems that you will actually encounter \cite{haldar2019applying}. (b) Quality-of-context bias: users make click decisions not only by the relevance of the clicked result, but also by the overall quality of the surrounding results in the list \cite{joachims2007evaluating}. The pairwise formulation measures the relative relevance by constructing pairs. (c) Intention bias: for example, hotels on the second page generally have higher CTR than those on the first page because users entering second pages are more engaged. A pointwise model tends to incorrectly favor the best hotels on the second page. 

% \begin{figure}[htp]
%   \centering
%   \includegraphics[width=12cm]{Bellagio}
%   \caption{Definition of commerce clicks and Hotel Review clicks. A commerce click brings user to our partner's site where user makes bookings, while a Hotel Review click brings user to our Hotel Review page. }
%   \label{fig:bellagio}
% \end{figure}

The training dataset of our search ranking model includes searches where at least one click/booking happens. We will be optimizing for bookings, similar to other OTAs \cite{grbovic2018real,haldar2019applying, bernardi2019150}, but will use clicks as supplemental data to facilitate the training process as we have far more clicks than bookings \cite{grbovic2018real,karmaker2017application}. 
% The definition of commerce clicks and Hotel Review clicks is shown in Figure \ref{fig:bellagio}. 
Moreover, optimizing for bookings to some extent addresses the concern of presentation bias. We use NDCG as our primary ranking metric, and the labels of hotel impressions are assigned based on the following rules:
\begin{itemize}
\item Booking: label = 5
\item Click into booking page: label = 2
\item Click into Hotel Review page: label = 1
\item No click: label = 0
\end{itemize}

To create personalization features, we trained hotel embeddings using word2vec with within-geo negative sampling, to account for congregated search (i.e. users frequently search only within a certain geo but not across geos), and these hotels embeddings were used to generate similarity features based on user behavior for real-time personalization similar to \cite{grbovic2018real}. We modified the gensim library \cite{rehurek_lrec} to allow this within-geo negative sampling. 

To reduce latency and better serve the online traffic, we use a two-stage ranking algorithm in our online ranking serving system, where an initial simple model retrieves a set of hotels (candidate generation) and a second complex model re-ranks them before presenting to users \cite{dang2013two}. This approach allows fast real-time recommendations and has been widely used in Google \cite{covington2016deep}, Facebook \cite{he2014practical} and Pinterest \cite{liu2017related,zhai2017visual}. 

\subsection{Experimental results}

To evaluate the effectiveness of propensity sampling, we ran three online experiments in Tripadvisor Hotels Best Value Sort for 2 weeks, and evaluated the performance in terms of clicks. For all the experiments, the propensity of results on or above LastClickedPos is kept as 1 (Equation \ref{eq: propensity above}), while two different variants on how to assign propensity to results below LastClickedPos will be tested against the control. 

\begin{itemize}
\item Control: 100\% sampling, $P(E=1 | k > LastClickedPos) = 100\%$
\item Test 1: 80\% sampling, $P(E=1 | k > LastClickedPos) = 80\%$
\item Test 2: propensity sampling, where $P(E=1 | k > LastClickedPos) = \frac{P(E=1 | k)}{P(E=1 | LastClickedPos)}$, and $P(E=1 | k)$ is calculated based on Section \ref{sec:Propensity Estimation}.
\end{itemize}

% The test result is shown in Table \ref{tab:Test Result}. Compared with the control model, the model with propensity sampling (Test 2) improves the commerce clicks by 1.8\% and Hotel Review clicks by 1.1\% (statistically significant). It also outperforms the model with a constant sampling rate of 80\% (Test 1), by 1.3\% in commerce clicks and 2.0\% in Hotel Review clicks (statistically significant). The model with 80\% sampling shows similar results as the control model, where the commerce clicks are brought up by 0.5\%, at the cost of 0.9\% fewer Hotel Review clicks. 

% \begin{table}
%  \caption{Online test results of different methods. Gain is relative to control. }
%   \centering
%   \begin{tabular}{c|c|c}
%     Name     & Commerce Click     & Hotel Review Click \\
%     \hline
%     Test 1: 80\% sampling & +0.5\%  & -0.9\%     \\
%     Test 2: propensity sampling & +1.8\%  & +1.1\%     \\
%   \end{tabular}
%   \label{tab:Test Result}
% \end{table}

The test result is shown in Table \ref{tab:Test Result}. Compared with the control model, the model with propensity sampling (Test 2) improves the clicks by 1.5\% (statistically significant). It also outperforms the model with a constant sampling rate of 80\% (Test 1), by 1.7\% in clicks (statistically significant). The model with 80\% sampling shows flat results compared with the control model.

\begin{table}
 \caption{Online test results of different methods. Gain is relative to control. }
  \centering
  \begin{tabular}{c|c|c}
    Name     &  Clicks  \\
    \hline
    Test 1: 80\% sampling & -0.2\%   \\
    Test 2: propensity sampling & +1.5\% \\
  \end{tabular}
  \label{tab:Test Result}
\end{table}

\section{Conclusion}
\label{sec:conclusion}

Although there is no widely accepted way of handling position bias in training LTR models, the importance of handling such bias should not be overlooked. In this work, we put forward a simple and easily adoptable method that fully leverages user actions with propensity sampling, and proves that it is effective through an online experiment. Online test results show that this method leads to significant performance improvements. Compared with large investments on infrastructures to support more complex models \cite{hazelwood2018applied}, this method requires minimal efforts without a higher level of model complexity, but is still able to improve search ranking significantly.

% \subsection{Figures}
% See Figure \ref{fig:fig1}. Here is how you add footnotes. \footnote{Sample of the first footnote.}

% \begin{figure}
%   \centering
%   \fbox{\rule[-.5cm]{4cm}{4cm} \rule[-.5cm]{4cm}{0cm}}
%   \caption{Sample figure caption.}
%   \label{fig:fig1}
% \end{figure}

% \subsection{Tables}
% \lipsum[12]
% See awesome Table~\ref{tab:table}.

% \begin{table}
%  \caption{Sample table title}
%   \centering
%   \begin{tabular}{lll}
%     \toprule
%     \multicolumn{2}{c}{Part}                   \\
%     \cmidrule(r){1-2}
%     Name     & Description     & Size ($\mu$m) \\
%     \midrule
%     Dendrite & Input terminal  & $\sim$100     \\
%     Axon     & Output terminal & $\sim$10      \\
%     Soma     & Cell body       & up to $10^6$  \\
%     \bottomrule
%   \end{tabular}
%   \label{tab:table}
% \end{table}

\bibliographystyle{unsrt}  
\bibliography{references}  %%% Remove comment to use the external .bib file (using bibtex).


% George Kour and Raid Saabne.
% \newblock Real-time segmentation of on-line handwritten arabic script.
% \newblock In {\em Frontiers in Handwriting Recognition (ICFHR), 2014 14th
%   International Conference on}, pages 417--422. IEEE, 2014.

% \end{thebibliography}

\end{document}